\documentclass[english,aps,twocolumn,pra]{revtex4}
\usepackage[T1]{fontenc}
\usepackage[latin9]{inputenc}
\usepackage{float}
\usepackage{amsmath}
\usepackage{amssymb}
\usepackage{graphicx}
\usepackage{esint}

\makeatletter
\@ifundefined{textcolor}{}
{%
 \definecolor{BLACK}{gray}{0}
 \definecolor{WHITE}{gray}{1}
 \definecolor{RED}{rgb}{1,0,0}
 \definecolor{GREEN}{rgb}{0,1,0}
 \definecolor{BLUE}{rgb}{0,0,1}
 \definecolor{CYAN}{cmyk}{1,0,0,0}
 \definecolor{MAGENTA}{cmyk}{0,1,0,0}
 \definecolor{YELLOW}{cmyk}{0,0,1,0}
 }

\makeatother

\usepackage{babel}
\begin{document}
%%%%%%%%%%%%%%%%%%%%%%%%%%%%%% Makra

\global\long\def\r{\left(\mathbf{r}\right)}

\global\long\def\rp{\left(\mathbf{r}'\right)}

\global\long\def\rt{\left(\mathbf{r}\tau\right)}

\global\long\def\rtp{\left(\mathbf{r}\tau'\right)}

\global\long\def\rpt{\left(\mathbf{r}'\tau\right)}

\global\long\def\rptp{\left(\mathbf{r}'\tau'\right)}

\global\long\def\Sh{\hat{\mathbf{S}}}

\global\long\def\rtrptp{\left(\mathbf{r}\tau;\mathbf{r}'\tau'\right)}

\global\long\def\dSh{\cdot\hat{\mathbf{S}}}

\global\long\def\n{\mathbf{n}}

\global\long\def\np{\dot{\mathbf{n}}}

\global\long\def\nnp{\mathbf{n}\cdot\dot{\mathbf{n}}}

\global\long\def\nknp{\mathbf{n}\times\dot{\mathbf{n}}}

\global\long\def\nh{\hat{n}}

\global\long\def\a{\alpha}

\global\long\def\b{\beta}

\global\long\def\s{\sigma}

\global\long\def\D{\Delta}

\global\long\def\sp{\sigma'}

\global\long\def\sb{\sigma"}

\global\long\def\Lb{{\bf \Lambda}}

\global\long\def\wm{\omega_{m}}

\global\long\def\wn{\omega_{n}}

\global\long\def\wl{\omega_{\ell}}

\global\long\def\wmp{\omega_{m'}}

\global\long\def\wnp{\omega_{n'}}

\global\long\def\kwm{\left(\mathbf{k}\omega_{m}\right)}

\global\long\def\uz{U_{0}}

\global\long\def\ud{U_{2}}

\global\long\def\d{\partial}

\global\long\def\e{\epsilon}

\global\long\def\mcS{\mathcal{S}}

\global\long\def\mcD{\mathcal{D}}

\global\long\def\mcH{\mathcal{H}}

\global\long\def\Ek{E_{\mathbf{k}}}

\global\long\def\zz{\left(\mathbf{0}0\right)}

%%%%%%%%%%%%%%%%%%%%%%%%%%%%%%%%%%%%

\title{Atom-atom correlations in time-of-flight imaging of ultra-cold bosons
in optical lattices}

\author{T. A. Zaleski}

\affiliation{Institute of Low Temperature and Structure Research, Polish Academy
of Sciences, POB 1410, 50-950 Wroc\l{}aw 2, Poland}

\author{T. K. Kope\'{c}}

\affiliation{Institute of Low Temperature and Structure Research, Polish Academy
of Sciences, POB 1410, 50-950 Wroc\l{}aw 2, Poland}
\begin{abstract}
We study the spatial correlations of strongly interacting bosons in
a ground state, confi{}ned in two-dimensional square and three-dimensional
cubic lattice. Using combined Bogoliubov method and the quantum rotor
approach, we map the Hamiltonian of strongly interacting bosons onto
U(1) phase action in order to calculate the atom-atom correlations
decay along the principal axis and a diagonal of the lattice plane
direction as a function of distance. Lower tunneling rates lead to
quicker decays of the correlations, which character becomes exponential.
Finally, correlation functions allow us to calculate quantities that
are directly bound to experimental outcomes, namely time-of-flight
absorption images and resulting visibility. Our results contain all
the characteristic features present in experimental data (transition
from Mott insulating blob to superfluid peaks, etc.), which emphasizes
the usability of the proposed approach. 
\end{abstract}


\maketitle

\section{Introduction}

During the last years enormous progress was made in the experimental
study of cold atoms in optical lattices \cite{optical_lattices}.
The great advantage of optical lattices as analog simulators of strongly
correlated Hamiltonians lies in the ability of optical lattices to
accurately implement lattice models without impurities or defects.
Furthermore, ultra-cold atoms confined in optical lattice structure
provide a very clean experimental realization of a strongly correlated
many-body problem \cite{quantum_many_body}. Strong correlation effects,
which imply enhanced quantum fluctuations, are playing an increasingly
important role in recent experiments on dilute quantum gases \cite{strong_correlations}.
To underline the importance of correlations and fluctuations involved
the studies of correlation functions are called for. One strategy
for boosting the importance of correlations and fluctuations involves
the study of the atomic correlators as a function of control of coupling
parameters. The atom-atom correlation function is defined by: 
\begin{equation}
C\left(\mathbf{r},\mathbf{r}'\right)=\left\langle a\left(\mathbf{r}\right)a^{\dagger}\left(\mathbf{r}'\right)\right\rangle ,\label{eq:correlation_function-generic}
\end{equation}
where $a\left(\mathbf{r}\right)$ is a bosonic operator and $\left\langle \dots\right\rangle $
is statistical averaging \cite{translation_satement}. This quantity
is required for computing time-of-flight absorption images, as obtained
with ultra-cold atoms released from optical lattices \cite{Gerbier,Hoffman}.
We will calculate this quantity using the quantum rotor approach developed
in our previous works \cite{polak1,zaleski1}, however now substantially
supplemented by the implementation of the Bogoliubov method \cite{bogoliubov}. 

In experiment, one can envisage to measure the momentum distribution
of the out-going atoms by taking a selective time of flight image
(TOF). For homogeneous case, time of flight measurements probe the
single-particle Green\textquoteright{}s function at equal times i.e.
the one-body density matrix. In an experiment of time-of-flight imaging,
the cloud of ultra-cold atoms is first suddenly released from the
harmonic trap. After a time of flight $t$, the position of the atoms
is proportional to the momentum of the atoms in the initial cloud.
Finally, an absorption image of the expanding cloud of atoms is taken
by a probing laser. The resulting image provides directly the distribution
of the momentum space $n(\mathbf{k})$. It is our goal of the presented
paper to calculate the time-of-flight patterns using the combined
Bogoliubov method and the quantum rotor approach and show that our
approach recreates all the characteristic features that are observed
in experimental settings. The plan of the paper is as follows: in
Section II, we introduce the microscopic Bose-Hubbard model relevant
for the description of strongly interacting bosons in an optical lattice.
In the next Section, we perform the evaluation of the atom-atom correlation
function by splitting the bosonic field into its amplitude and phase.
Furthermore, in Section III, by implementing the Bogoliubov method
to the amplitude part of the bosonic field and combining its outcome
with the quantum rotor approach, we derive analytically the explicit
expressions for the atom-atom correlation function, time-of-flight
absorption images and visibility. In Section IV, results of our calculations
are plotted: the spatial dependence of the correlation function as
a function of lattice distance $\mathbf{R}$, time-of-flight images
and visibility for various model parameters. Finally, we conclude
in the Section V.

\section{Model Hamiltonian}

From a theoretical point of view, description of the bosons in optical
lattice can be achieved through the definition of a microscopic Hamiltonian,
which can capture the main physics of these systems: the Bose Hubbard
Hamiltonian. Within this model, the bosons move on a lattice within
a tight-binding scheme and correlation is introduced through an on-site
repulsive term, since in real Bose gases the interaction between atoms
cannot be neglected in the physical description of the gas. We consider
a second quantized, bosonic Hubbard Hamiltonian in the form \cite{bemodel1,bemodel2}:

\begin{eqnarray}
\mathcal{H} & = & -t\sum_{\left\langle \mathbf{r},\mathbf{r}'\right\rangle }\left[a^{\dagger}\left(\mathbf{r}\right)a\left(\mathbf{r}'\right)+a^{\dagger}\left(\mathbf{r}'\right)a\left(\mathbf{r}\right)\right]\nonumber \\
 &  & +\frac{U}{2}\sum_{\mathbf{r}}n^{2}\r-\overline{\mu}\sum_{\mathbf{r}}n\left(\mathbf{r}\right).\label{eq:mainham}
\end{eqnarray}
 The constant $t$ represents nearest neighbors tunneling matrix element
and is responsible for the dynamical hopping of bosons from one optical
lattice site to another. During a jump between two neighboring sites
a boson gains energy $t$. The constant $U$ is the strength of the
on-site repulsive interaction of bosons. Adding a boson to already
occupied site costs energy $U$. Furthermore, $\overline{\mu}=\mu+\frac{U}{2}$,
where $\mu$ is a chemical potential controlling the average number
of bosons. The operators $a^{\dagger}\left(\mathbf{r}\right)$ and
$a\left(\mathbf{r}'\right)$ create and annihilate bosons on sites
$\mathbf{r}$ and $\mathbf{r}'$ of a regular two-dimensional (2D)
lattice with the nearest neighbors hopping denoted by summation over
$\left\langle \mathbf{r},\mathbf{r}'\right\rangle $. A total number
of sites is equal to $N$ and the boson number operator $n\r=a^{\dagger}\left(\mathbf{r}\right)a\r$.
We use the Hamiltonian in Eq. (\ref{eq:mainham}) to describe a homogeneous
(translationally invariant) system, omitting the effect of the external
magnetic potential that is usually superimposed on top of the optical
lattice potential in order to additionally trap the atoms. The external
potential can be included in the Hamiltonian as $\sum_{\mathbf{r}}\epsilon\r n\r$
and would couple to the chemical potential term. We discuss such a
scenario and its consequences in the Section IV.\textbf{ }The realization
of the Bose-Hubbard Hamiltonian using optical lattices has the advantage
that the interaction matrix element $U$ and the tunneling matrix
element $t$ can be controlled by adjusting the intensity of the laser
beams. The Hamiltonian and its descendants have been widely studied
within the last years. The phase diagram and ground-state properties
include the mean-fi{}eld ansatz \cite{key-1}, strong coupling expansions
\cite{key-2,key-3,key-4}, the quantum rotor approach \cite{key-5},
methods using the density matrix renormalization group DMRG \cite{key-6,key-7,key-8,key-9},
and quantum Monte Carlo QMC simulations \cite{key-10,key-11,key-12,key-13}.

\section{Atomic correlation functions}

\subsection{Transformation to the amplitude-phase variables}

The statistical sum of the system defined by Eq. \eqref{eq:mainham}
can be written in a path integral form with use of complex fields,
$a\rt$ depending on the {}``imaginary time'' $0\le\tau\le\beta\equiv1/k_{B}T$,
(with $T$ being the temperature) that satisfy the periodic condition
$a\rt=a(\mathbf{r}\tau+\beta)$: 
\begin{equation}
Z=\int\left[\mathcal{D}\overline{a}\mathcal{D}a\right]e^{-\mathcal{S}\left[\overline{a},a\right]},\label{statsum}
\end{equation}
 where the action $\mcS$ is equal to: 
\begin{equation}
\mathcal{S}\left[\overline{a},a\right]=\int_{0}^{\beta}d\tau\left[\mathcal{H}\left(\tau\right)+\sum_{\mathbf{r}}\overline{a}\left(\mathbf{r}\tau\right)\frac{\partial}{\partial\tau}a\left(\mathbf{r}\tau\right)\right].
\end{equation}
Now, we are briefly introducing the quantum rotor approach, which
has already been employed for the calculation of the phase diagram
of the cold bosons in optical lattice \cite{polak1}. The fourth-order
term in the Hamiltonian in Eq. (\ref{eq:mainham}) can be decoupled
using the Hubbard-Stratonovich transformation with an auxiliary field
$V\rt$: 
\begin{eqnarray}
 &  & e^{-\frac{U}{2}\sum_{\mathbf{r}}\int_{0}^{\beta}d\tau n^{2}\left(\mathbf{r}\tau\right)}\nonumber \\
 &  & \,\,\,\,\,\propto\int\frac{\mathcal{D}V}{\sqrt{2\pi}}e^{\sum_{\mathbf{r}}\int_{0}^{\beta}d\tau\left[-\frac{V^{2}\left(\mathbf{r}\tau\right)}{2U}+iV\left(\mathbf{r}\tau\right)n\left(\mathbf{r}\tau\right)\right]}.
\end{eqnarray}
 The fluctuating {}``imaginary chemical potential'' $iV\rt$ can
be written as a sum of static $V^{\circ}\left(\mathbf{r}\right)$
and periodic function: 
\begin{eqnarray}
V\left(\mathbf{r}\tau\right) & = & V^{\circ}\left(\mathbf{r}\right)+\tilde{V}\left(\mathbf{r}\tau\right),
\end{eqnarray}
 where, using Fourier series: 
\begin{eqnarray}
\tilde{V}\left(\mathbf{r}\tau\right) & = & \frac{1}{\beta}\sum_{\ell=1}^{\infty}\tilde{V}\left(\mathbf{r}\omega_{\ell}\right)\left(e^{i\omega_{\ell}\tau}+e^{-i\omega_{\ell}\tau}\right),
\end{eqnarray}
 with the Bose-Matsubara frequencies are $\omega_{\ell}=2\pi\ell/\beta$
and $\ell=0,\pm1,\pm2,\dots$. Introducing the U(1) phase field $\phi\rt$
via the Josephson-type relation \cite{kopec1}: 
\begin{equation}
\dot{\phi}\left(\mathbf{r}\tau\right)=\tilde{V}\left(\mathbf{r}\tau\right)
\end{equation}
with $\dot{\phi}\rt=\d\phi\rt/\d\tau$ we can now perform a local
gauge transformation to new bosonic variables:

\begin{equation}
a\rt=b\rt e^{i\phi\rt}.\label{eq:variable_transform}
\end{equation}
As a result, the strongly correlated bosonic system is transformed
into a weakly interacting bosons, submerged into the bath of strongly
fluctuating gauge potentials (which interactions are governed by the
high energy scale of $U$). The order parameter is defined by: 
\begin{equation}
{\bf \Psi}_{B}=\left\langle a\rt\right\rangle =\left\langle b\rt\right\rangle \psi_{B}.\label{eq:largepsi_b-1}
\end{equation}
However, one should note that a nonzero value of the amplitude $\left\langle b\rt\right\rangle $
is not sufficient for superfluidity. To achieve this, also the phase
(rotor) variables, must become stiff and coherent. The phase order
parameter is defined by: 
\begin{equation}
\psi_{B}=\left\langle e^{i\phi\rt}\right\rangle _{\phi},\label{eq:phase_order_parameter-1}
\end{equation}
where $\left\langle \dots\right\rangle _{\phi}$ is averaging over
phase action to be calculated in the next subsection. This reflects
the fact that all the atoms in the condensate have the same phase
and form a coherent matter wave. Thus the condensate possess a well
defined phase associated with the concept of so-called spontaneously
broken U(1) gauge symmetry. 

After the variable transformations the statistical sum becomes: 
\begin{equation}
Z=\int\left[\mcD\bar{b}\mcD b\right]\left[\mcD\phi\right]e^{-\mcS\left[\bar{b},b,\phi\right]}
\end{equation}
 with the action: 
\begin{eqnarray}
 &  & \mathcal{S}\left[\overline{b},b,\phi\right]=\sum_{\mathbf{r}}\int_{0}^{\beta}d\tau\overline{b}\left(\mathbf{r}\tau\right)\frac{\partial}{\partial\tau}b\left(\mathbf{r}\tau\right)\nonumber \\
 &  & -t\sum_{\left\langle \mathbf{r},\mathbf{r}'\right\rangle }\int_{0}^{\beta}d\tau\left\{ e^{-i\left[\phi\rt-\phi\rpt\right]}\overline{b}\left(\mathbf{r}\tau\right)b\left(\mathbf{r}'\tau\right)+h.c.\right\} \nonumber \\
 &  & +\sum_{\mathbf{r}}\int_{0}^{\beta}d\tau\left[U\left\langle \overline{b}\left(\mathbf{r}\tau\right)b\left(\mathbf{r}\tau\right)\right\rangle _{b}-\overline{\mu}\right]\overline{b}\left(\mathbf{r}\tau\right)b\left(\mathbf{r}\tau\right)\nonumber \\
 &  & +\sum_{\mathbf{r}}\int_{0}^{\beta}d\tau\left[\frac{1}{2U}\dot{\phi}^{2}\rt+i\frac{\overline{\mu}}{U}\dot{\phi}\rt\right],
\end{eqnarray}
which will be used as a departure point for obtaining bosonic and
phase-only actions.

\subsection{Transformation to the rotor representation for phase variables}

The statistical sum can be integrated over the phase or bosonic variables
with the phase or bosonic action: 
\begin{eqnarray}
\mcS_{\phi}\left[\phi\right] & = & -\ln\int\left[\mcD\bar{b}\mcD b\right]e^{-\mcS\left[\bar{b},b,\phi\right]},\nonumber \\
\mathcal{S}_{b}\left[\overline{b},b\right] & = & -\ln\int\left[\mcD\phi\right]e^{-\mcS\left[\bar{b},b,\phi\right]}\label{eq:phase_bosonic_action_generic}
\end{eqnarray}
to obtain:
\begin{equation}
Z=\int\left[\mcD\phi\right]e^{-\mcS_{\phi}\left[\phi\right]}=\int\left[\mcD\overline{b}\mathcal{D}b\right]e^{-\mcS_{b}\left[\overline{b},b\right]}.\label{eq:phase_partition_function}
\end{equation}
In order to calculate the phase-only action, we use the following
approximation:
\begin{equation}
a\rt=b\rt e^{i\phi\rt}\approx b_{0}e^{i\phi\rt},
\end{equation}
where $b_{0}$ is static bosonic amplitude, which will be calculated
in the next Section. The phase-only action from Eq. (\ref{eq:phase_bosonic_action_generic})
can be written explicitly: 
\begin{eqnarray}
\mcS_{\phi}\left[\phi\right] & = & \int_{0}^{\beta}d\tau\left\{ \sum_{\mathbf{r}}\left[\frac{\dot{\phi}^{2}\rt}{2U}+i\frac{\overline{\mu}}{U}\dot{\phi}\rt\right]\right.\nonumber \\
 & - & \left.J\sum_{\left\langle \mathbf{r},\mathbf{r}'\right\rangle }\cos\left[\phi\rt-\phi\rpt\right]\right\} ,\label{eq:phase_action}
\end{eqnarray}
 where $J=t\left|b_{0}\right|^{2}$ represents the stiffness for the
phase field. It is clear that the phase action is non-linear in the
phase variables $\phi\rt$ and the statistical sum in Eq. (\ref{eq:phase_partition_function})
cannot be calculated exactly. Because of the trigonometric nature
of the phase variables it is useful to introduce a new uni-modular
collective field $z\rt=e^{i\phi\rt}$, which will be treated within
the quantum rotor approach by making use of the following resolution
of unity: 
\begin{equation}
1\equiv\int d\overline{z}dz\delta\left[z\rt-e^{i\phi\rt}\right]\delta\left[\overline{z}\rt-e^{-i\phi\rt}\right].\label{eq:resolution_unity}
\end{equation}
The unit length constraint for $z\rt$ variables is implemented on
average by the formula (see, Ref. \onlinecite{kopec2}):

\begin{equation}
\delta\left[N-\sum_{\mathbf{r}}\overline{z}\rt z\rt\right]=\int d\lambda e^{N\lambda-\lambda\sum_{\mathbf{r}}\overline{z}\rt z\rt},
\end{equation}
which introduces a Lagrange multiplier $\lambda$. Substituting Eq.
\eqref{eq:resolution_unity} into Eq. \eqref{eq:phase_partition_function}
and integrating by the cumulant expansion over the phase variables,
the partition function reads: 
\begin{eqnarray}
Z & = & \int\left[\mathcal{D}\overline{z}\mathcal{D}z\right]d\lambda e^{N\lambda-\mathcal{S}_{z}\left[\overline{z},z\right]},\label{eq:partition_function_zz}
\end{eqnarray}
In the thermodynamic limit ($N\rightarrow\infty$), the integral \eqref{eq:partition_function_zz}
can be performed exactly by the saddle-point method. The quantum rotor
action: 
\begin{eqnarray}
\mathcal{S}_{z}\left[\overline{z},z\right] & = & \sum_{\left\langle \mathbf{r},\mathbf{r}'\right\rangle }\int_{0}^{\beta}d\tau d\tau'\left[\left(\lambda\delta_{\mathbf{r}\mathbf{r}'}-tb_{0}^{2}\right)\delta\left(\tau-\tau'\right)\right.\nonumber \\
 &  & \left.+\delta_{\mathbf{r}\mathbf{r}'}K^{-1}\left(\tau-\tau'\right)\right]\overline{z}\rt z\rptp,\label{eq:rotor-action}
\end{eqnarray}
 where $K^{-1}\left(\tau-\tau'\right)$ is a phase correlator, which
includes dynamic effect of the U(1) phase field: 
\begin{equation}
K^{-1}\left(\tau-\tau'\right)=\left\langle e^{i\phi\rt-i\phi\rtp}\right\rangle _{0}.\label{eq:phase_correlator}
\end{equation}
Furthermore, the average 
\begin{equation}
\left\langle \dots\right\rangle _{0}=\frac{\int\left[\mcD\phi\right]\dots e^{-\mcS_{0}\left[\phi\right]}}{\int\left[\mcD\phi\right]e^{-\mcS_{0}\left[\phi\right]}}
\end{equation}
 is taken only over non-interacting quantum rotors: 
\begin{equation}
\mathcal{S}_{0}=\sum_{\mathbf{r}}\int_{0}^{\beta}d\tau\left[\frac{\dot{\phi}^{2}\rt}{2U}+i\frac{\overline{\mu}}{U}\dot{\phi}\rt\right].
\end{equation}
 The Fourier transform of the correlator in Eq. (\ref{eq:phase_correlator})
in zero temperature limit reads: 
\begin{equation}
K^{-1}\left(\wl\right)=\frac{U}{4}-U\left[v\left(\frac{\overline{\mu}}{U}\right)+\frac{i\wl}{U}\right]^{2},
\end{equation}
where $v\left(x\right)=x-\left[x\right]-\frac{1}{2}$, and $\left[x\right]$
is the floor function, which gives the greatest integer less then
or equal to $x$. Non-zero value of the order parameter $\Psi_{B}$
signals a bosonic condensation that we identify as superfluid state.
The phase order parameter in the quantum rotor model can be written
as: 
\begin{equation}
1-\psi_{B}^{2}=\frac{1}{N}\sum_{\mathbf{r}}\int_{0}^{\beta}d\tau\left\langle \overline{z}\rt z\rt\right\rangle \label{eq:convergence_constraint}
\end{equation}
which fixes value of the phase order parameter $\psi_{B}$. The Lagrange
multiplier saddle-point value {}``sticks\textquotedblright{} at criticality
to the value $\lambda_{0}$ given by 
\begin{equation}
\lambda_{0}-J\left(\mathbf{k}=0\right)+K^{-1}\left(\omega_{\ell=0}\right)=0,\label{eq:l0_criticality}
\end{equation}
and obeys the Eq. \eqref{eq:l0_criticality} in the whole low temperature
ordered phase.

\subsection{Bogoliubov transformation of the amplitude bosonic variables}

In the Bogoliubov approximation the complicated many-body quartic
Hamiltonian is reduced to a quadratic one, which can be diagonalized
exactly. In order to calculate the bosonic action in Eq. (\ref{eq:phase_bosonic_action_generic}),
we follow the Bogoliubov approach by splitting the bosonic operator
into a Bose condensate macroscopic occupation $N_{0}=b_{0}^{2}$ and
non-condensed fluctuation part $b_{d}\left(\mathbf{r}\right)$ \cite{bogoliubov}.
As a consequence, the original operator splits into a sum and according
to Eq. \eqref{eq:variable_transform} at an alternative representation:
\begin{equation}
b\rt=b_{0}+b_{d}\rt.\label{eq:ab0bd}
\end{equation}
Substituting the Eq. \eqref{eq:ab0bd} into the action in Eq. \eqref{eq:phase_bosonic_action_generic},
and neglecting the terms of the order higher than two in $b_{d}\r$
operators, the action reads: 
\begin{equation}
\mathcal{S}=\mathcal{S}_{0}+\mathcal{S}_{1}+\mathcal{S}_{2},\label{eq:action_bogoliubov}
\end{equation}
where the zero, first and the second order terms containing interactions
within and between both sub-systems are: 
\begin{eqnarray}
\mathcal{S}_{0} & = & N\beta\left[-tz-\overline{\mu}+\frac{U}{2}\left|b_{0}\right|^{2}\right]\overline{b}_{0}b_{0}.
\end{eqnarray}
The linear part in the fluctuation operators reads: 
\begin{eqnarray}
 &  & \mathcal{S}_{1}=\sum_{\mathbf{r}}\int_{0}^{\beta}d\tau\left\{ \left[-zt-\overline{\mu}+U\left|b_{0}\right|^{2}\right]\overline{b}_{0}b_{d}\rt\right\} \nonumber \\
 &  & +\sum_{\mathbf{r}}\int_{0}^{\beta}d\tau\left\{ \left[-zt-\overline{\mu}+U\left|b_{0}\right|^{2}\right]b_{0}\overline{b}_{d}\rt\right\} ,
\end{eqnarray}
and finally, the quadratic part: 
\begin{eqnarray}
 &  & \mathcal{S}_{2}=\frac{1}{2}\sum_{\mathbf{r}}\int_{0}^{\beta}d\tau d\tau'\,\left[\begin{array}{cc}
\overline{b}_{d}\rt, & b_{d}\rtp\end{array}\right]\nonumber \\
 &  & \times\left[\begin{array}{cc}
2U\left|b_{0}\right|^{2}-A_{+} & U\left|b_{0}\right|^{2}\\
U\left|b_{0}\right|^{2} & 2U\left|b_{0}\right|^{2}-A_{-}
\end{array}\right]\left[\begin{array}{c}
b_{d}\rt\\
\overline{b}_{d}\rtp
\end{array}\right]\nonumber \\
 &  & -t\sum_{\left\langle \mathbf{r}>\mathbf{r}'\right\rangle }\int_{0}^{\beta}d\tau\,\left[\overline{b}_{d}\rt b_{d}\rpt+h.c.\right],
\end{eqnarray}
where:
\begin{equation}
A_{\pm}\left(\tau-\tau'\right)=\overline{\mu}\mp\delta\left(\tau-\tau'\right)\frac{\partial}{\partial\tau}.
\end{equation}
 Furthermore, the value of $b_{0}$ amplitude can be calculated in
the saddle point: 
\begin{equation}
\frac{\d\mathcal{S}\left[b_{0}\right]}{\d b_{0}}=-tz-\overline{\mu}+U\left|b_{0}\right|^{2}=0,
\end{equation}
 which results in: 
\begin{equation}
\left|b_{0}\right|^{2}=z\frac{t}{U}+\frac{\overline{\mu}}{U}.
\end{equation}
 This implies that the linear term in Eq. (\ref{eq:action_bogoliubov})
is $\mathcal{S}_{1}=0$ and the bosonic action is given by: 
\begin{equation}
\mathcal{S}\left[\overline{b},b\right]=\mathcal{S}_{0}+\mathcal{S}_{2}.\label{eq:bosonic_action}
\end{equation}

\subsection{Correlation function}

Summarizing up the results of the preceding sections, the correlation
function in Eq. \eqref{eq:correlation_function-generic} can be written
as a product of two correlation functions of amplitude and rotor fields:
\begin{eqnarray}
C\left(\mathbf{R}\right) & \equiv & C_{z}\left(\mathbf{R}\right)C_{b}\left(\mathbf{R}\right),\label{eq:CorrelationFunction}
\end{eqnarray}
 where 
\begin{eqnarray}
C_{z}\left(\mathbf{R}\right) & \equiv & C_{z}\left(\mathbf{r}\tau;\mathbf{r}'\tau\right)=\left\langle z\rt\overline{z}\rpt\right\rangle _{z}\nonumber \\
C_{b}\left(\mathbf{R}\right) & \equiv & C_{b}\left(\mathbf{r}\tau;\mathbf{r}'\tau\right)=\left\langle b\left(\mathbf{r}\right)\overline{b}\left(\mathbf{r}'\right)\right\rangle _{b}.\label{eq:CzCb}
\end{eqnarray}
The averagings appearing in Eq. (\ref{eq:CzCb}) are defined by: 
\begin{eqnarray}
\left\langle \dots\right\rangle _{z} & = & \frac{\int\left[\mcD\overline{z}\mcD z\right]\dots e^{-\mcS_{z}\left[\overline{z},z\right]}}{\int\left[\mcD\overline{z}\mcD z\right]e^{-\mcS_{z}\left[\overline{z},z\right]}}\nonumber \\
\left\langle \dots\right\rangle _{b} & = & \frac{\int\left[\mcD\bar{b}\mcD b\right]\dots e^{-\mcS_{b}\left[\overline{b},b\right]}}{\int\left[\mcD\bar{b}\mcD b\right]e^{-\mcS_{b}\left[\overline{b},b\right]}},
\end{eqnarray}
 where $\mcS_{z}\left[\overline{z},z\right]$ and $\mcS_{b}\left[\overline{b},b\right]$
are given in Eqs. (\ref{eq:rotor-action}) and (\ref{eq:bosonic_action}),
respectively. As a result, the correlation function in the real space
splits into a product of averages from bosonic and phase sectors.
The Green's function in the bosonic sector reads: 
\begin{equation}
G_{b}\kwm=b_{0}^{2}+G_{bd}^{22}\kwm,
\end{equation}
 where: 
\begin{eqnarray}
 &  & G_{bd}\kwm\nonumber \\
 &  & =\left[\begin{array}{cc}
\frac{2t\left(\varepsilon_{\mathbf{0}}-\varepsilon_{\mathbf{k}}\right)+U\left|b_{0}\right|^{2}-i\wl}{E_{\mathbf{k}}^{2}-\left(i\wl\right)^{2}} & -\frac{U\left|b_{0}\right|^{2}}{E_{\mathbf{k}}^{2}-\left(i\wl\right)^{2}}\\
-\frac{U\left|b_{0}\right|^{2}}{E_{\mathbf{k}}^{2}-\left(i\wl\right)^{2}} & \frac{2t\left(\varepsilon_{\mathbf{0}}-\varepsilon_{\mathbf{k}}\right)+U\left|b_{0}\right|^{2}+i\wl}{E_{\mathbf{k}}^{2}-\left(i\wl\right)^{2}}
\end{array}\right],\,\,\,\,\,\,\,\,\,\,
\end{eqnarray}
 and 
\begin{equation}
E_{\mathbf{k}}=\sqrt{2t\left(\frac{z}{2}-\varepsilon_{\mathbf{k}}\right)\left[2t\left(\frac{z}{2}-\varepsilon_{\mathbf{k}}\right)+2U\left|b_{0}\right|^{2}\right]}
\end{equation}
with a dispersion for a simple cubic lattice: 
\begin{equation}
\varepsilon_{\mathbf{k}}=\cos\left(ak_{x}\right)+\cos\left(ak_{y}\right)+\cos\left(ak_{z}\right).\label{eq:dispersion}
\end{equation}
 On the other hand, the phase sector leads to the phase Green's function:
\begin{eqnarray}
G_{\phi}\kwm & = & \left\langle z\kwm\overline{z}\kwm\right\rangle \nonumber \\
 & = & \frac{1}{\lambda_{0}-2tb_{0}^{2}\varepsilon_{\mathbf{k}}+K^{-1}\left(\wm\right)}.\label{eq:Gfi}
\end{eqnarray}
 As a result, the correlation function finally reads: 
\begin{equation}
C\left(\mathbf{R}\right)=G_{\phi}\left(\mathbf{R}\right)\left[\left|b_{0}\right|^{2}+G_{bd}\left(\mathbf{R}\right)\right].\label{eq:correlation_function}
\end{equation}

\subsection{Spatial dependence of the correlation functions}

The correlation function in Eq. \eqref{eq:correlation_function-generic}
can be calculated along the lattice axis ($x$ or $y$), or along
the diagonal of the square lattice:
\begin{eqnarray}
\mathbf{R}_{\mathrm{parallel}} & = & \left(na_{x},0\right),\nonumber \\
\mathbf{R}_{\mathrm{diagonal}} & = & \left(na_{x},na_{y}\right),
\end{eqnarray}
where $a_{x}$, and $a_{y}$ are lattice constants in $x$, and $y$
direction, respectively and $n$ is an integer. The results for two-dimensional
square lattice are presented in Figs. \ref{fig:Correlation-function}a
and b for the Mott state within the first and the fifth Mott lobe,
respectively. The correlation function is plotted in logarithmic scale.
As expected, lower tunneling rates lead to quicker decays of the correlations.
Also, the character of the decay becomes exponential as on-site repulsion
$U$ gets stronger (dependence becomes linear in logarithmic scale).
Correlations calculated along the lattice diagonal are weaker than
those parallel to the lattice axis (see, Fig. \ref{fig:Correlation-function}c),
which is also in agreement with results from Ref. \onlinecite{correlations}). 

From this result, we can deduce that the asymptotic spatial behavior
of the correlation function $C\left(\mathbf{R}\right)$ is given by
the formula:
\begin{equation}
C\left(\mathbf{R}\right)=\exp\left(-\frac{\left|\mathbf{R}\right|}{\xi}\right),
\end{equation}
where $\xi$ stands for the correlation length. On the other hand,
from Eq. (\ref{eq:correlation_function}) we see that the critical
behavior (associated with ordering of the phase variables) is governed
by $G_{\phi}\left(\mathbf{R}\right)$ {[}see, Eq. (\ref{eq:Gfi}){]}
-- a correlation function of the quantum spherical model \cite{criticalexponents},
for which a universal quantum-critical ($T=0$) properties imply that
the correlation length as a function of on-site scattering $U$, close
to the critical point, should behave as: 
\begin{equation}
\xi=\left(\frac{U-U_{c}}{U_{c}}\right)^{-1},
\end{equation}
where $U_{c}$ is the critical value of $U$.

\begin{center}
\begin{figure}[H]
\centering{}\includegraphics[scale=0.5]{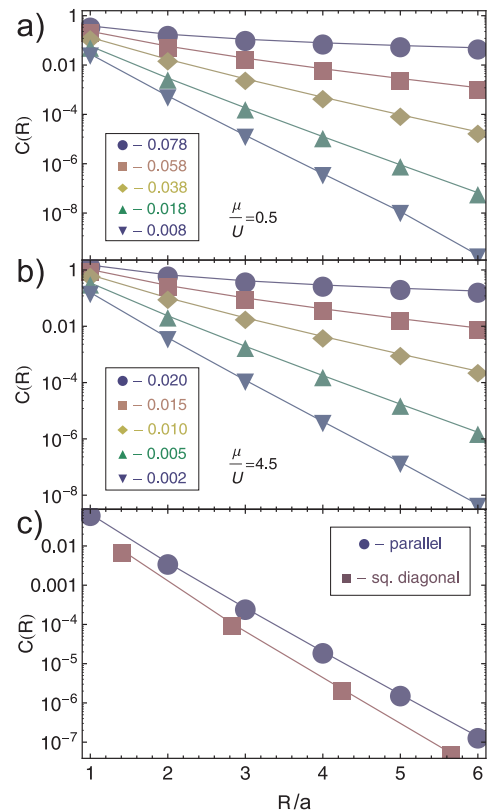}\caption{(Color online) Correlation function $C\left(\mathbf{R}\right)$ along
axis direction for various values of $\frac{t}{U}$ for two-dimensional
lattice inside a) the first and b) the fifth Mott lobe; c) comparison
of correlations decay for $\frac{t}{U}=0.02$ and $\frac{\mu}{U}=0.5$
along axis (circles) and lattice in-plane diagonal (squares) for two-dimensional
lattice. \label{fig:Correlation-function}}
\end{figure}

\par\end{center}

\section{Time-of-flight patterns}

The phase coherence properties can be estimated from time of flight
images. When atoms in the superfluid state are released from the trap
and optical lattice, phase coherence leads to an interference pattern
with interference peaks arranged in the order reflecting the optical
lattice symmetry. However, the Mott insulating phase does not have
long range phase coherence and does not show the interference pattern
in the time-of-flight images. To identify the ordered state, we concentrate
on the momentum distribution of particle number $n\left(\mathbf{k}\right)$,
a quantity of basic interest that encodes the strong correlations
of the system:

\begin{equation}
n\left(\mathbf{k}\right)=\sum_{\mathbf{R}}C\left(\mathbf{R}\right)e^{i\mathbf{k}\mathbf{R}}.
\end{equation}
This quantity is measured in time-of-flight experiments, which are
performed by releasing the atomic cloud from the optical trap potential.
Due to the fact that the expansion is mainly ballistic, the momentum
dependence of particle numbers represents the interference pattern,
which typically has two distinct types of behavior: a wide maximum
signifying the Mott insulating incoherent state, and a sharp peaks
located in the reciprocal lattice vectors, which are considered as
a signature of superfluidity in the system. According to the formulas
in Eqs. (\ref{eq:CorrelationFunction})-(\ref{eq:Gfi}), we have:
\begin{eqnarray}
n\left(\mathbf{k}\right) & = & \left\langle a_{\mathbf{k}}^{\dagger}\left(\tau\right)a_{\mathbf{k}}\left(\tau\right)\right\rangle \nonumber \\
 & = & m_{0}\frac{1}{\beta}\sum_{m}G_{bd}\left(\mathbf{k}\wm\right)+b_{0}\frac{1}{\beta}\sum_{m}G_{\phi d}\left(\mathbf{k}\wm\right)\nonumber \\
 &  & \,\,\,\,+\frac{1}{N\beta^{2}}\sum_{\mathbf{k}',m,m'}G_{bd}\left(\mathbf{k}'\wm\right)G_{\phi d}\left(\mathbf{k}-\mathbf{k}';\wmp\right)\nonumber \\
 & = & m_{0}G_{bd}\left(\mathbf{k}\right)+b_{0}G_{\phi d}\left(\mathbf{k}\right)\nonumber \\
 &  & \,\,\,\,+\frac{1}{N}\sum_{\mathbf{k}'}G_{bd}\left(\mathbf{k}'\right)G_{\phi d}\left(\mathbf{k}-\mathbf{k}'\right),\label{eq:nkasgf}
\end{eqnarray}
where:
\begin{eqnarray}
G_{bd}\left(\mathbf{k}\right) & = & \frac{1}{2}\left[-1+\frac{2t\left(\varepsilon_{0}-\varepsilon_{\mathbf{k}}\right)+n_{0}U}{E_{\mathbf{k}}}\coth\frac{\beta E_{\mathbf{k}}}{2}\right]\nonumber \\
G_{\phi d}\left(\mathbf{k}\right) & = & \frac{\coth\left\{ \frac{\beta U}{2}\left[\Omega_{\mathbf{k}}+v\left(\frac{\overline{\mu}}{U}\right)\right]\right\} }{4\Omega_{\mathbf{k}}}\nonumber \\
 & + & \frac{\coth\left\{ \frac{\beta U}{2}\left[\Omega_{\mathbf{k}}-v\left(\frac{\overline{\mu}}{U}\right)\right]\right\} }{4\Omega_{\mathbf{k}}},
\end{eqnarray}
with:

\begin{equation}
\Omega_{\mathbf{k}}=\sqrt{2\frac{tb_{0}^{2}}{U}\left(\varepsilon_{0}-\varepsilon_{\mathbf{k}}\right)+v^{2}\left(\frac{\overline{\mu}}{U}\right)+\frac{\delta\lambda}{U}}.
\end{equation}

We now turn to the description of the interference pattern observed
after release of the atom cloud from the optical lattice and a period
of free expansion, where the phase coherence of the atoms on the optical
lattice can be directly probed. The density distribution of the expanding
cloud after time $t$ can be represented as follows \cite{tof1,tof2,tof3}:

\begin{equation}
n\left(\mathbf{r}\right)=\left.\left(\frac{m}{\hbar t}\right)^{3}\left|W\left(\mathbf{k}\right)\right|^{2}n\left(\mathbf{k}\right)\right|_{\mathbf{k}=\frac{m\mathbf{r}}{\hbar t}},\label{eq:nr}
\end{equation}
where $m$ is the atomic mass and $t$ is cloud expansion time. The
quantity $W\left(\mathbf{k}\right)$ in the Eq. \ref{eq:nr} is the
Fourier transform of the Wannier function in the lowest Bloch band.
Typically, the trapping potential is well approximated by a harmonic
function so that the envelope has the Gaussian form:
\begin{equation}
\left|W\left(\frac{m\mathbf{r}}{\hbar t}\right)\right|^{2}\approx\frac{1}{\pi^{3/2}w_{t}}\exp\left(-\frac{\mathbf{r}^{2}}{w_{t}^{2}}\right),\label{eq:wannierenvelope}
\end{equation}
where $w_{t}=\hbar t/mw_{0}$ with $w_{0}$ being the size of the
on-site Wannier function. Therefore, in order to compare the interference
pattern with experiments, we have to calculate $n\left(\mathbf{r}\right)$.
However, the experimental absorption pictures are two-dimensional
projections of the three-dimensional optical lattice, i.e. they are
column-integrated momentum distribution over one axis:
\begin{equation}
n_{\perp}\left(k_{x},k_{y}\right)=\int dk_{z}\left|W\left(\mathbf{k}\right)\right|^{2}n\left(\mathbf{k}\right).
\end{equation}
Observing that $n\left(\mathbf{k}\right)=n\left[G_{xd}\left(\mathbf{k}\right)\right]$,
where $x$=$\phi,b$ (see, Eq. \ref{eq:nkasgf}) and making an expansion:
\begin{equation}
G_{xd}\left(\mathbf{k}\right)=g_{0x}+g_{x}\varepsilon\left(\mathbf{k}\right)+O\left[\varepsilon\left(\mathbf{k}\right)\right],
\end{equation}
where $\varepsilon\left(\mathbf{k}\right)$ is given by Eq. (\ref{eq:dispersion}),
we see that the contribution:
\begin{equation}
\int dk_{z}\left|W\left(\mathbf{k}\right)\right|^{2}\cos ak_{z}=\sqrt{\pi}w_{0}e^{-\left(\frac{aw_{0}}{2}\right)^{2}}\approx0
\end{equation}
is negligible for sufficiently large optical lattice depths, especially
deep in the Mott regime. Thus, we can consider an effective two-dimensional
system a good approximation of the three-dimensional one. The result
is presented in Fig. \ref{fig:Simulation-of-time-of-flight}. In the
superfluid phase, the sharp peaks emerge denoting long-range phase
coherence. In the Mott phase, the momentum distribution becomes a
broad, featureless maximum. 

To asses the interference pattern, a useful quantity is often introduced,
which is called visibility and measured in many experiments \cite{vis1,vis2,vis3}:
\begin{equation}
v=\frac{n_{max}-n_{min}}{n_{max}+n_{min}}.
\end{equation}
The minimum and maximum intensities $n_{min}$ and $n_{max}$ are
measured at the same distance from the cloud center, which for two-dimensional
square lattice are at $\mathbf{k}=\left(2\pi/\sqrt{2}a,2\pi/\sqrt{2}a\right)$
and $\mathbf{k}=\left(2\pi/a,0\right)$, respectively, at which the
Wannier envelope in Eq. (\ref{eq:wannierenvelope}) cancels out. The
resulting dependence of visibility on interaction ratio is presented
in Fig. \ref{fig:Visibility}. In the superfluid phase, its value
is almost 1 signaling strong coherence peaks in time-of-flight patters.
Deeper within Mott lobes, the visibility is decreasing, while the
time-of-flight pattern becomes a circularly symmetric blob. However,
in Fig. \ref{fig:Simulation-of-time-of-flight} the phase transition
occurs abruptly and signature of superfluidity, namely sharp peaks,
disappear fast within the Mott phase. This is not usually observed
in experiments, where the remainders of the peaks are visible even
for systems deep in the Mott phase. It results from the fact that
in the experimental setups the atoms are trapped not only within the
optical potential, but additionally within magnetic parabolic trap
that is symmetric around the trap center. As a result, the system
is no longer homogeneous and correlation functions become dependent
not only on the distance, but also on specific location, which makes
it impossible to calculate them in analytic way anymore. The best
we can do is to assume that the homogeneous solution works locally,
depending on a {}``local'' chemical potential $\overline{\mu}\r=\overline{\mu}+\epsilon\r$
(see, Ref. \cite{bloch_review}). Since, the time-of-flight patterns
are global absorption images of atoms released from the traps, we
sum them over each lattice site {[}which has a local value of $\overline{\mu}\r${]}
multiplying by a local number of atoms. Because, the value of $\overline{\mu}\r$
is dependent on the distance from the trap center (decreasing from
its maximal value to zero, on the trap boundary), for chosen interaction
ratio $t/U$, parts of the system are superfluid, while the other
parts can be in the Mott regime. As a results, even if the majority
of the system is in the Mott state, some remnant signatures of superfluidity
may still be visible.\textbf{ }In order to make a reliable comparison
of our theoretical prediction with experiments, we need a translation
of the parameters of the Bose-Hubbard model $t/U$ into the quantities
determining the experimental setup, namely $V_{0}/E_{R}$ ($V_{0}$
is the potential depth and $E_{R}$ is the recoil energy). The bandwidth
parameter\textbf{ $t$} is essentially the gain in kinetic energy
due to nearest neighbor tunneling. In the limit $V_{0}\gg E_{R}$
it can be obtained from the exact result for the width of the lowest
band in the 1D Mathieu-equation \cite{theortoexp}: 
\begin{equation}
t\approx\frac{4}{\sqrt{\pi}}\left(\frac{V_{0}}{E_{R}}\right)^{3/4}e^{-2\sqrt{V_{0}/E_{R}}}.
\end{equation}
The relevant interaction parameter $U$ is thus given by an integral
over the on-site wave function $W\left(\mathbf{x}\right)$ via:
\begin{equation}
U=\frac{4\pi\hbar^{2}a}{m}\int\left|W\left(\mathbf{x}\right)\right|^{4}\approx4\sqrt{2\pi}\frac{a}{\lambda}\left(\frac{V_{0}}{E_{R}}\right)^{3/4},
\end{equation}
where, $\lambda$ and $a$ are laser wave length and scattering length,
respectively (for $^{87}\mbox{Rb}$, $a_{s}=5.45\mbox{nm}$ and $\lambda=850\mbox{nm}$).
It follows, that:
\begin{equation}
\frac{t}{U}=\frac{1}{\sqrt{2}\pi}\left(\frac{\lambda}{a}\right)e^{-2\sqrt{V_{0}/E_{R}}}.
\end{equation}
After averaging of the time-of-flight images over a range of chemical
potential $\mu$, we can present our results in a form that is directly
comparable with experimental results in Fig. \ref{fig:Comparison-of-tof}
(see, the figure caption). It is clear that all the characteristic
features have been reproduced. Also, because of the better resolution
and lack of noise in the theoretical results, hints of superfluid
state that is present in narrow spheres around the trap center, are
visible even for deep lattice $V_{0}/E_{R}=20$. 

\begin{center}
\begin{figure}[H]
\begin{centering}
\includegraphics[scale=0.4]{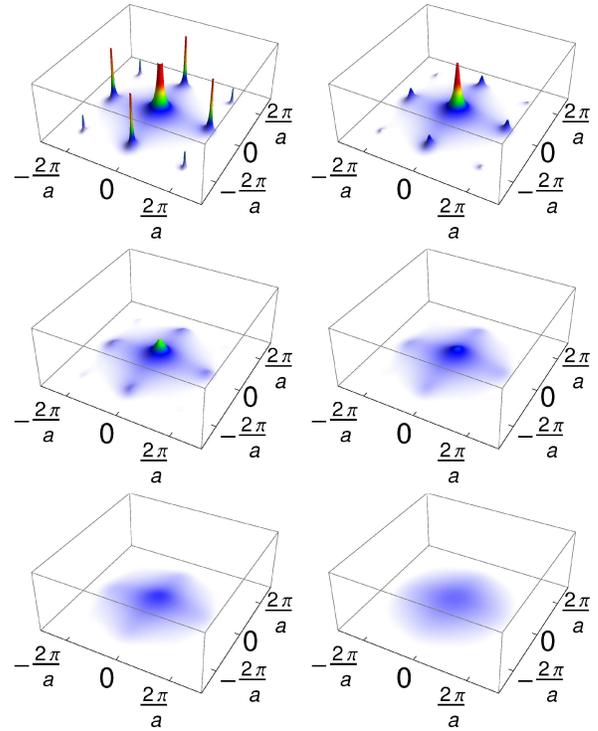}
\par\end{centering}

\caption{(Color online) Simulation of time-of-flight absorption images {[}from.
Eq. \eqref{eq:nr}{]} for various interactions strength $t/U$: transition
from superfluid (top-left, $t/U=0.085$) to Mott phase (top-right
to the bottom-right $t/U=0.085,\,0.07,\,0.055,\,0.04,\,0.025,\,0.01$,
respectively), for $\mu/U=0.5$.\label{fig:Simulation-of-time-of-flight}}
\end{figure}

\par\end{center}

\begin{center}
\begin{figure}[H]
\begin{centering}
\includegraphics[scale=0.5]{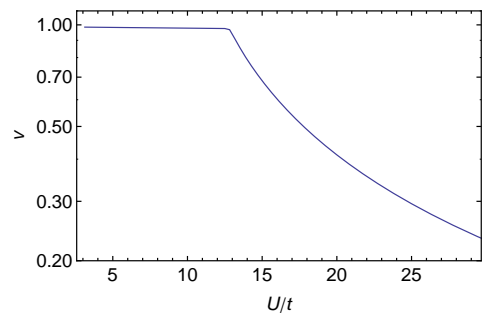}
\par\end{centering}

\caption{(Color online) Visibility calculated from simulated time-of-flight
images as a function of coupling ratio $U/t$ for $\mu/U=0.5$.\label{fig:Visibility}}
\end{figure}

\par\end{center}

\begin{center}
\begin{figure}[H]
\begin{centering}
\includegraphics[scale=0.4]{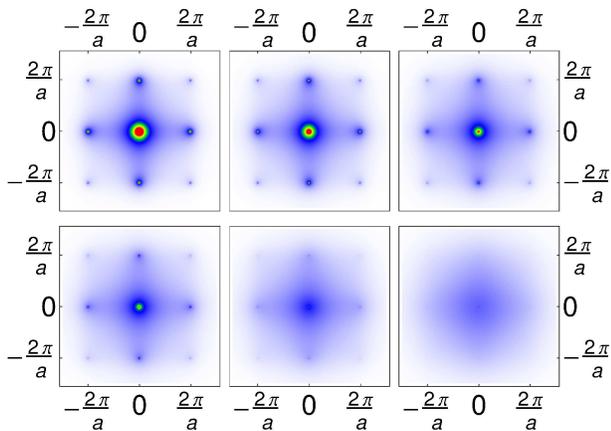}
\par\end{centering}

\caption{(Color online) Simulated time-of-flight patterns for comparison with
experimental time-of-flight absorption images (for values of $V_{0}/E_{R}=7,10,13,14,17,20$,
around 150,000 lattice sites and average number of up to 2.5 atoms
per lattice site in the center of the trap, chosen as in Ref. {[}\onlinecite{optical_lattices}{]},
Fig. 2c-h, ).\label{fig:Comparison-of-tof}}
\end{figure}

\par\end{center}

\section{Conclusions}

In summary, we have studied the correlations of cold atoms loaded
in a two dimensional optical lattice. In the superfluid regime the
validity of the Bogoliubov approximation is restricted to the very
weakly interacting regime ($U/t\ll1$) and its breakdown as quantum
correlations become important. In general Bogoliubov quasi-particle
states correspond to solutions of the approximate Hamiltonian with
a plane-wave character. As quantum correlations become important,
the exact eigenstates of the interacting system do not necessarily
have the simple plane wave character, especially in the commensurate
filling situation where for a critical value of $U/t$ the system
exhibits the superfuid-Mott insulator transition. To address these
issues a new theory beyond the simple Bogoliubov approximation was
developed that incorporates the phase degrees of freedom via the quantum
rotor approach to describe regimes beyond the very weakly interacting
one. This scenario provided a picture of quasiparticles and energy
excitations in the strong interaction limit, where the transition
between the superfluid and the Mott state is be driven by phase fluctuations.
Taking advantage of the macroscopically populated condensate state,
we have separated the problem into the amplitude of the Bose field
and the fluctuating phase that was absent in the original Bogoliubov
problem. Subsequently, the functional formulation this formalism was
shown to be a powerful tool that incorporates properly the interaction
aspects characteristic of the quantum phase dynamics. This formalism
provides a useful framework, where the one particle the correlation
functions are treated self-consistently and permits us to test and
simulate Bose-Hubbard Hamiltonian with a whole range of phenomena.
To explore these a diagnostics tools that include spectroscopy of
occupation numbers and correlation measurement is called for. Especially,
the time of flight imaging is a very powerful technique to probe the
quantum gases in optical lattices. From the beginning of the ultra-cold
atom field, it permitted to observe the Bose Einstein condensation.
In this work we have calculated time-of-flight patterns and the relevant
correlation functions on two dimensional optical lattice and compared
them with experimental data. Our theoretical pictures match the experimental
ones, where one can observe the formation of a central and various
neighboring peaks, out of featureless Mott state, which subsequently
become sharper in the superfluid phase as a result of a phase coherence.
It would be interesting to map out the nature of the Bose transition
in optical lattice systems by using further diagnostic tools such
as Bragg spectroscopy, which reveals the whole momentum structure
of the single particle correlation function. A detailed understanding
of this quantity is a challenging task in the further experimental
studies of trapped ultra-cold gases. 
\begin{acknowledgments}
We would like to acknowledge support from Polish Ministry of Science
and Higher Education (Grant No. N N202 045537).\end{acknowledgments}


\begin{thebibliography}{References}
\bibitem{translation_satement}In our calculations, we assume that
the system is homogeneous, i.e. $C\left(\mathbf{R}\right)\equiv C(\mathbf{r}-\mathbf{r}')$,
with a lattice vector $\mathbf{R}=\mathbf{r}-\mathbf{r}'$. Since,
the system is invariant under translations by integer multiples of
lattice vectors (homogeneous), $C\left(\mathbf{R}\right)$ depends
only on $\mathbf{R}$, but not on the individual sites $\mathbf{r}$
and $\mathbf{r}'$.
\end{thebibliography}
\end{document}